\documentclass[iop]{emulateapj}

\usepackage{color}
\usepackage{xcolor}
\usepackage{gensymb}
\usepackage{booktabs}

\newcommand\rmxaa{Rev. Mexicana Astron. Astrofis.}

\shorttitle{Survival of interstellar molecules}
\shortauthors{Hincelin, Wakelam, Commer\c con et al. }

\begin{document}

\title{Survival of interstellar molecules to prestellar dense core collapse and early phases of disk formation}

\author{U. Hincelin}
\affil{Department of Chemistry, University of Virginia, Charlottesville, VA 22904, USA}
\email{ugo.hincelin@virginia.edu}

\author{V. Wakelam}
\affil{Univ. Bordeaux, LAB, UMR 5804, F-33270, Floirac, France and \\
CNRS, LAB, UMR 5804, F-33270, Floirac, France}

\author{B. Commer\c con}
\affil{Laboratoire de radioastronomie, LERMA, Observatoire de Paris, Ecole Normale Sup\'{e}rieure (UMR 8112 CNRS), 24
rue Lhomond, 75231 Paris Cedex 05, France}

\author{F. Hersant and S. Guilloteau}
\affil{Univ. Bordeaux, LAB, UMR 5804, F-33270, Floirac, France and \\
CNRS, LAB, UMR 5804, F-33270, Floirac, France}

\begin{abstract}
{An outstanding question} of astrobiology is the link between the chemical composition {of planets, comets, and other solar system bodies} and the molecules formed in the interstellar medium.
Understanding the chemical and physical evolution of the matter leading to the formation of protoplanetary disks is an important step for this.
{We provide some new clues to this long-standing problem using three-dimensional chemical simulations of the early phases of disk formation: we interfaced the full gas-grain chemical model Nautilus with the radiation-magnetohydrodynamic model {\ttfamily{RAMSES}}, for different configurations and intensities of magnetic field.}
{Our results show that the chemical content (gas and ices) is globally conserved during the collapsing process, from the parent molecular cloud to the young disk surrounding the first Larson core.}
{A qualitative comparison with cometary composition suggests that comets are constituted of different phases, some molecules being direct tracers of interstellar chemistry, while others, including complex molecules, seem to have been formed in disks, where higher densities and temperatures allow for an active grain surface chemistry.
The latter phase, and its connection with the formation of the first Larson core, remains to be modeled.}

\end{abstract}


\section{Introduction}
\label{intro}
{Low mass stars like our Sun are formed by the collapse of a prestellar dense core.
The matter falls in the center of the core, creating a protoplanetary disk surrounding a protostar.
Planets, comets and {other planetary/circumstellar bodies} can then be formed in the disk.
The chemical composition of the interstellar matter and its evolution during the formation of the disk are important to better understand the formation process of these planets and other bodies.
In order to study the chemical evolution during the early phases of Solar System formation, one needs to know the evolution of the density and the temperature of the matter during the collapse process.
Indeed, the chemical composition and its evolution depend mostly on these quantities.
Besides, the chemical evolution is a non-equilibrium process, and thus it is necessary to follow the different phases of the collapse and disk formation to correctly deal with this subject.}

{The first phases are the following ones.
After the beginning of the collapse, a first hydrostatic core at equilibrium is reached (when the density exceeds $10^{-13}$~g.cm$^{-3}$), which is called the first Larson core \citep{1969MNRAS.145..271L}.
This is the first step to the formation of the star and disk system.
The accretion on the core continues, and when H$_2$ is dissociated (at a temperature of 2000~K), a second collapse begins, which leads to the formation of the second Larson core (when the density exceeds 1~g.cm$^{-3}$).
The second Larson core is the protostar.}
{First Larson cores have not been observed yet, mainly because they are embedded (high optical thickness) and have a short lifetime ($\sim1000$~yrs).
A few first candidates have been reported in recent years \citep{2006A&A...454L..51B,2010ApJ...715.1344C,2010ApJ...722L..33E,2011ApJ...742....1D,2011ApJ...743..201P,2012ApJ...751...89C,2012A&A...547A..54P} but their confirmation remains controversial since the expected observable signatures of first hydrostatic cores are still uncertain.}
{However, recent progress has been made to characterize dust emission of first Larson cores \citep{2012A&A...548A..39C}.}

{To understand the link between the chemical composition of the interstellar medium and the one observed in disks, comets and other Solar System bodies, the chemical evolution of the matter needs to be coupled to the physics of the prestellar collapse.}
The chemistry of collapsing cores {has been studied} using one-dimensional models \citep{1996ApJ...471..400C,2003ApJ...585..355R,2004A&A...418.1021D,2004ApJ...617..360L,2006A&A...457..927G,2008ApJ...674..984A,2008ApJ...682..283G}.
\cite{2009A&A...497..773V} and \cite{2011A&A...534A.132V} presented two-dimensional chemical evolution during the collapse of a molecular cloud to form a low-mass protostar and its surrounding disk.
{As a first step to study the detailed coupling between the dynamics of the formation of disks and their chemical composition, \citet{2012ApJ...758...86F} (hereafter FA12) performed three dimensional radiation-hydrodynamical simulations of star formation coupled with a detailed chemical network.}

{To go a step further, we present in this paper three-dimensional chemical simulations of the early formation of a primitive protoplanetary disk using a radiation-magneto-hydrodynamic model, where the magnetic field allows for the formation of more realistic disks, less prone to early fragmentation \citep{2011ApJ...742L...9C}.}
We have computed the evolution of the chemical composition of gas and ices using a full gas-grain chemical network, in a collapsing prestellar dense core {up to the first Larson core (just before the onset of the second collapse)}.
The core physical structure has been consistently derived using state-of-the-art radiation-magneto-hydrodynamic (RMHD) models of star formation.
From these simulations, we infer how the chemical composition of the initial parent cloud is modified by the formation of the disk.

{The paper is organized as follows.}
In section 2, the chemical and physical models are presented.
{The results are then shown and discussed respectively in sections 3 and 4.}

\section{Model}
\label{model}

For this modeling, we have used the gas-grain chemical code Nautilus \citep{2009A&A...493L..49H} with the physical conditions of a collapsing dense core computed using the adaptive mesh refinement code {\ttfamily{RAMSES}} \citep{2002A&A...385..337T}.

\subsection{The physical model {\ttfamily{RAMSES}}}

The physical structure of the collapsing core is derived using the RMHD solver of {\ttfamily{RAMSES}} which integrates the equations of ideal magnetohydrodynamics \citep{2006A&A...457..371F} and the equations of radiation-hydrodynamics under the grey flux-limited diffusion (FLD)  approximation \citep{2011A&A...529A..35C}.
The initial conditions are those used in \citet{2012A&A...545A..98C} and consist in a 1~M$_\odot$ sphere of uniform density $\rho_0=3.97\times 10 ^{-18}$~g.cm$^{-3}$ {(corresponding to a molecular hydrogen number density $n_{\rm H_2}\approx~10^6$~cm$^{-3}$)} and temperature $T_0=11$~K.
The sphere (radius $r_0\approx 3300$~AU) is in solid body rotation, and threaded by a uniform magnetic field. 
The strength of the magnetic field is expressed in terms of the mass-to-flux to critical mass-to-flux ratio
$\mu=(M_0/\Phi)/(M_0/\Phi)_\mathrm{c}$.
$M_0$ is equal to M$_\odot$.
$\Phi$ is the magnetic flux.
If $(M_0/\Phi)$ is lower than $(M_0/\Phi)_\mathrm{c}$ (i.e. $\mu$ is lower than 1), then the core is supported against its own gravity by the magnetic field, and does not collapse.
{In this study, we use two magnetization levels: a moderate one ($\mu=10$) and a low one ($\mu=200$).}
{For $\mu=10$, we use two different initial configurations: an angle $\Theta=0\degree$ between magnetic field lines and the rotation axis of the sphere, and an other angle $\Theta=45\degree$.}
{These values form three different models that we present in this paper: MU10$\Theta$0, MU10$\Theta$45 and MU200$\Theta$0 (see Table~\ref{tab:model3D_liste_modeles}).}
We computed the evolution of the collapsing dense core throughout the first collapse and first Larson core lifetime \citep{1969MNRAS.145..271L}, {i.e. $\sim 38\times 10^3$~yr}.
{ The RMHD method used in {\ttfamily{RAMSES}} has been well tested, as well as the numerical convergence by \citet{Commercon_2010}. Accordingly, the grid is dynamically refined as a function of the local Jeans length so that it is always resolved by at least 10 cells. In addition, it has been shown in previous studies that the grey FLD approximation is well adapted to study the early phases of star formation  by comparing to more advanced moment and multifrequency models \citep{2011A&A...530A..13C,2012A&A...543A..60}. }

\begin{table}[h]
\begin{center}
\caption{{Magnetic fields strength and orientations used for the current simulations.} \label{tab:model3D_liste_modeles}}
\begin{tabular}{ccc}
\tableline
\tableline
Model & $\mu$ & $\Theta$ (\degree) \\
\tableline
MU10$\Theta$0  & 10 (moderate) & 0 \\
MU10$\Theta$45 & 10 (moderate) & 45 \\
MU200$\Theta$0 & 200 (low) & 0 \\
\tableline
\end{tabular}
\end{center}
\end{table}

\subsection{The chemical model Nautilus}
\label{subsec:Nautilus}

Nautilus computes the abundance of species in the gas-phase and at the surface of the grains as a function of time, taking into account pure gas-phase chemistry (bimolecular reactions, dissociative and radiative recombinations and associations, electron attachments, dissociations and ionizations induced by cosmic-rays and UV photons...), interaction between the species in the gas-phase and the grain surfaces (adsorption, thermal desorption, and non-thermal desorption by cosmic-rays), and grain surface reactions (chemical reactions, and dissociation by UV photons and cosmic ray induced photons).
A more detailed description of this model and the chemical network can be found in  \citet{2010A&A...522A..42S}, and \citet{2011A&A...530A..61H}.
The network has been updated according to the latest recommendations from the KIDA\footnote{\samepage
KInetic Database for Astrochemistry, \url{http://kida.obs.u-bordeaux1.fr}, \cite{2012ApJS..199...21W}
}
experts until October 2011.
An electronic version of our network is available at \url{http://kida.obs.u-bordeaux1.fr/models}.

To compute the initial chemical composition of the sphere, we run Nautilus for dense cloud conditions (10~K, H density\footnote{\samepage
Most cloud simulations use a value of $2\times 10^4$~cm$^{-3}$.
Our higher value accounts for the fact that the matter is in the densest region of molecular cloud (where a dense core is formed).}
of $2\times 10^5$~cm$^{-3}$, cosmic-ray ionization rate of $1.3\times 10^{-17}$~s$^{-1}$ and a visual extinction of 30) for a time $t\sim 6\times 10^5$~yr, assuming that species are initially in the atomic form except for hydrogen, which is already molecular.
We used the elemental abundances from \citet{2011A&A...530A..61H}, with an oxygen elemental abundance relative to total hydrogen equal to $1.5\times 10^{-4}$ (C/O ratio equal to $1.13$).
For each element, we define the elemental abundance as the ratio of the number of nuclei both in the gas and on the dust grains to the total number of H nuclei.
This excludes the nuclei locked in the refractory part of the grains, that is why we did not use solar elemental abundances which takes into account this refractory part.
The chosen time $t$ corresponds to the maximum of agreement between observations of molecular clouds and our simulations (see Fig.3. of \cite{2011A&A...530A..61H}).
Using this chemical composition as the initial condition, we then run the chemistry during the collapse using the method described in the next subsection.

   \subsection{Interface between Nautilus and {\ttfamily{RAMSES}} codes}
   \label{interface}

\begin{figure}
\includegraphics[width=1.0\linewidth]{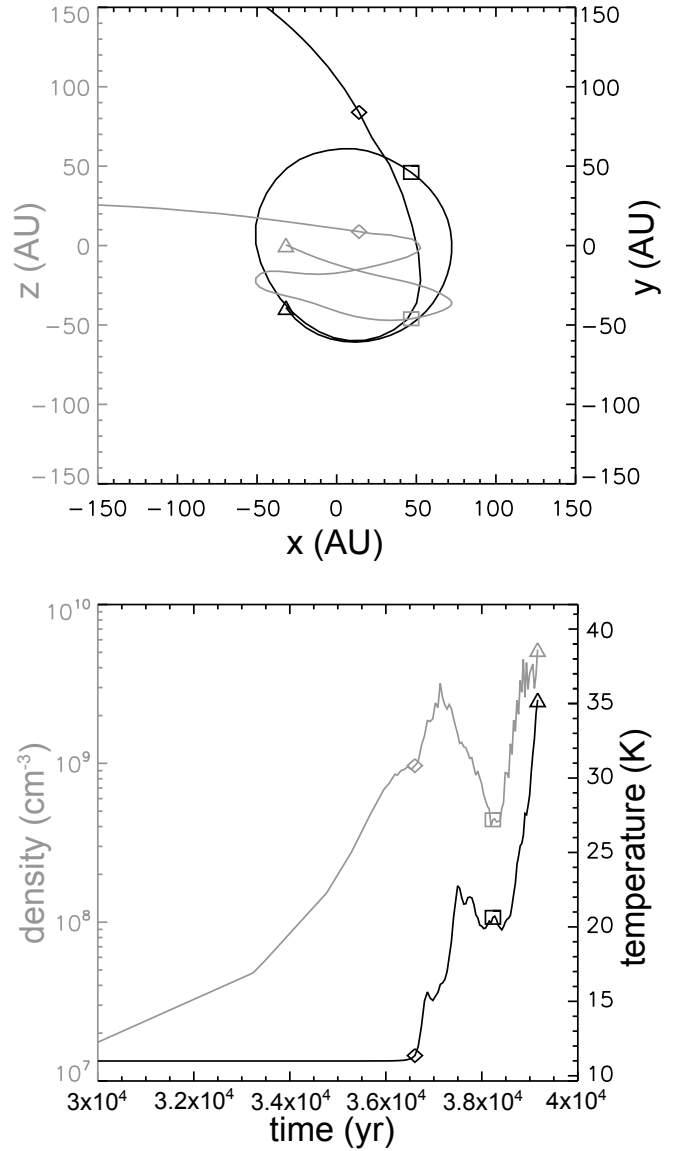}
\caption{{Example of a trajectory in cartesian coordinate system (top) and the associated temperature T(t) and density n(t) as a function of time t (bottom), for one given particle {of model MU10$\Theta$45}.
Diamond, square and triangle show positions (top) at specific times (bottom).
The center of first Larson core has coordinates $\rm (x,y,z)=(0,0,0)$.}\label{fig:trajectory_example}}
\end{figure}

{To simplify our simulations, already very time consuming, there is no feedback of the chemical composition calculated by Nautilus on the dynamics computed using the {\ttfamily{RAMSES}} code, i.e. Nautilus computations are a post-processing of {\ttfamily{RAMSES}} {simulations} (see section~\ref{sec:limits}).
To compute the chemistry, Nautilus needs two input data: the temperature and the density as a function of time. For both the physical and chemical calculations, we assume that the object is embedded in a cloud so that direct photoprocesses can be neglected {(visual extinction is fixed to 30)} and that the cosmic-ray ionization rate is constant {(fixed to $1.3\times 10^{-17}$~s$^{-1}$)}. The first step of the calculation is to obtain the time dependent physical conditions.
To this aim,} we introduce $10^6$ ``probe'' particles, initially uniformly distributed in the sphere, which handle the temperature, density and velocity history of the fluid through a Lagrangian approach.
{For any particle, at each time step, the position is registered, the temperature, density and velocity are extracted according {to} the local fluid properties, and the position at the next time step is iterated. 
At the end of this process, we obtain the positions, the velocities, and the couples temperature/density $\rm (T(t),n(t))$ as a function of time $t$ for the million particles of the model.
Fig.~\ref{fig:trajectory_example} shows an example of computed trajectory and its corresponding $\rm (T(t),n(t))$.}
Starting from the initial chemical composition of the sphere, the chemical evolution is then computed in 0D for each parcel of material (``probe'' particle) using the physical conditions derived from the RMHD simulations {(i.e. the $10^6$ couples $\rm (T(t),n(t))$)}.
Note that the chemistry does not have the time to reach steady state in any of our calculations.
This process gives the full 3D chemical structure of the collapsing dense core as a function of time.
Central Processing Unit (CPU) time to compute the chemical evolution of $10^6$ particles (i.e. for one model) is roughly $10^4$ hours (using JADE cluster from CINES).

\section{Results}

\subsection{Physical properties of disks}
Within this structure, we identify several components representing 1) the core itself (the first Larson core), 2) the outflow, 3) the rotating disk, 4) the magnetically supported pseudodisk, and 5) the envelope, based on criteria from \citet{2012A&A...543A.128J}.
More details on the extraction method will be given in a future paper (Hincelin et al. in preparation). 
{The rotating disk component} that we consider here is made by particles that satisfy the four following criteria:
\begin{enumerate}
\item an azimuthal velocity $v_\phi$ more than two times larger than their radial velocity $v_r$,
\item an azimuthal velocity more than two times larger than their vertical velocity $v_z$,
\item a rotational support (density of rotational kinetic energy $E_k=0.5\rho v_\phi^2$, where $\rho$ is the mass density) at least two times larger than the thermal support (density of internal energy $E_i=1/(\gamma -1)\times(\rho k_B T)/(\mu_{m} m_{\rm H})$, where $\gamma$, $k_B$, $\mu_m$, and $m_{\rm H}$ are respectively the adiabatic index, the Boltzmann constant, the mean molecular weight, and the mass of hydrogen nucleus),
\item and a density above $10^9$~cm$^{-3}$.
\end{enumerate}

{Fig.~\ref{fig:cartes_T} and \ref{fig:cartes_n} present selected particles that belong to the rotating disk for the three models presented in Table~\ref{tab:model3D_liste_modeles}.
The other components (central core, outflow, pseudodisk and envelope) will be discussed in a future paper (Hincelin et al. in preparation).}
{The plotted information (in color) are respectively the temperature and the density of the particles.}
{The disks are relatively cold, around 10~K, except at low radius where the temperature can reach about 100-200~K.
Density spreads from $10^9$ to $10^{11-12}$~cm$^{-3}$ from the outer edge of disks to their center.
The temperature profile has roughly a spherical symmetry (except for model MU200$\Theta$0 with the presence of four "hot spots" around the central core) which is due to the central core (see also Fig.2 in \cite{2012A&A...545A..98C}), while the density profile has a more complex shape, for example with the presence of spiral arms for $\mu = 10$ (MU10$\Theta$0 and MU10$\Theta$45 models).
{In these simulations, the disk of model MU10$\Theta$0 is about 200 AU in diameter, whereas the one of model MU10$\Theta$45 is roughly two times larger ($\sim 400$~AU).}
{Due to the low magnetization level, the disk of model MU200$\Theta$0 shows a more complex shape with fragmentation.
Four ``hot spots'' are identified around a central disk.
Each spot is characterized by an increase of both temperature and density.
The central disk is about 80~A.U. in diameter.}
{Table~\ref{tab:model3D_measurements} shows the range of sizes, temperatures and densities found in the disks at the end of simulations for the three models. One notes that the disk of model MU200$\Theta$0 is fragmented due to the low magnetization level, and the disk of model MU10$\Theta$45 is twisted because of the inclination between rotational axes and initial magnetic field lines.}

\begin{figure}
\includegraphics[width=1.0\linewidth]{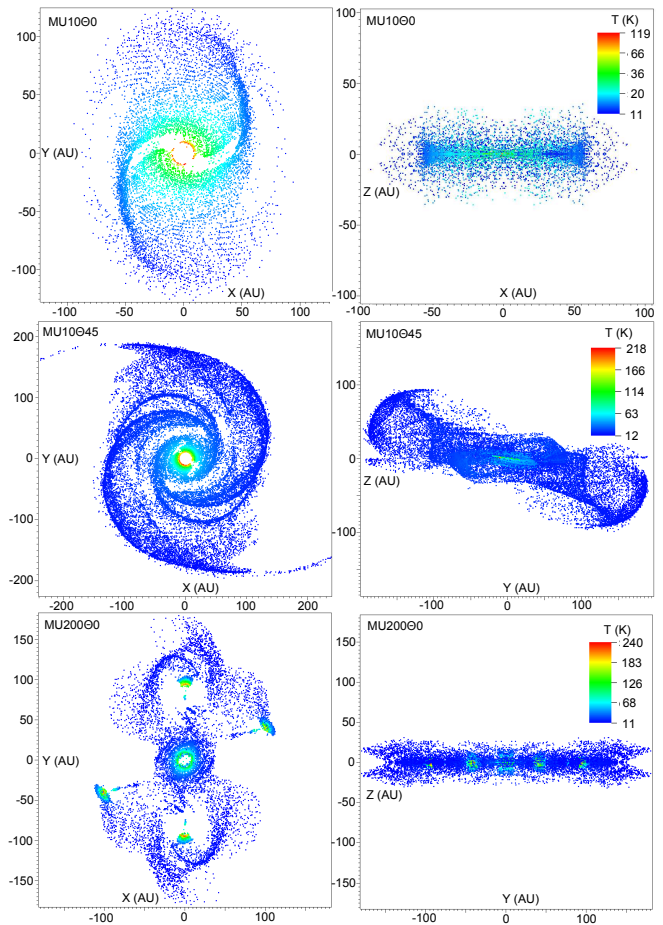}
\caption{{Temperature (K) in the disk for the three models MU10$\Theta$0, MU10$\Theta$45 and MU200$\Theta$0.
Face-on view on the left and edge-on view on the right.
Particles are projected in the plane (xOy), (xOz) or (yOz).
{The color coding} is linear.}}
\label{fig:cartes_T}
\end{figure}

\begin{figure}
\includegraphics[width=1.0\linewidth]{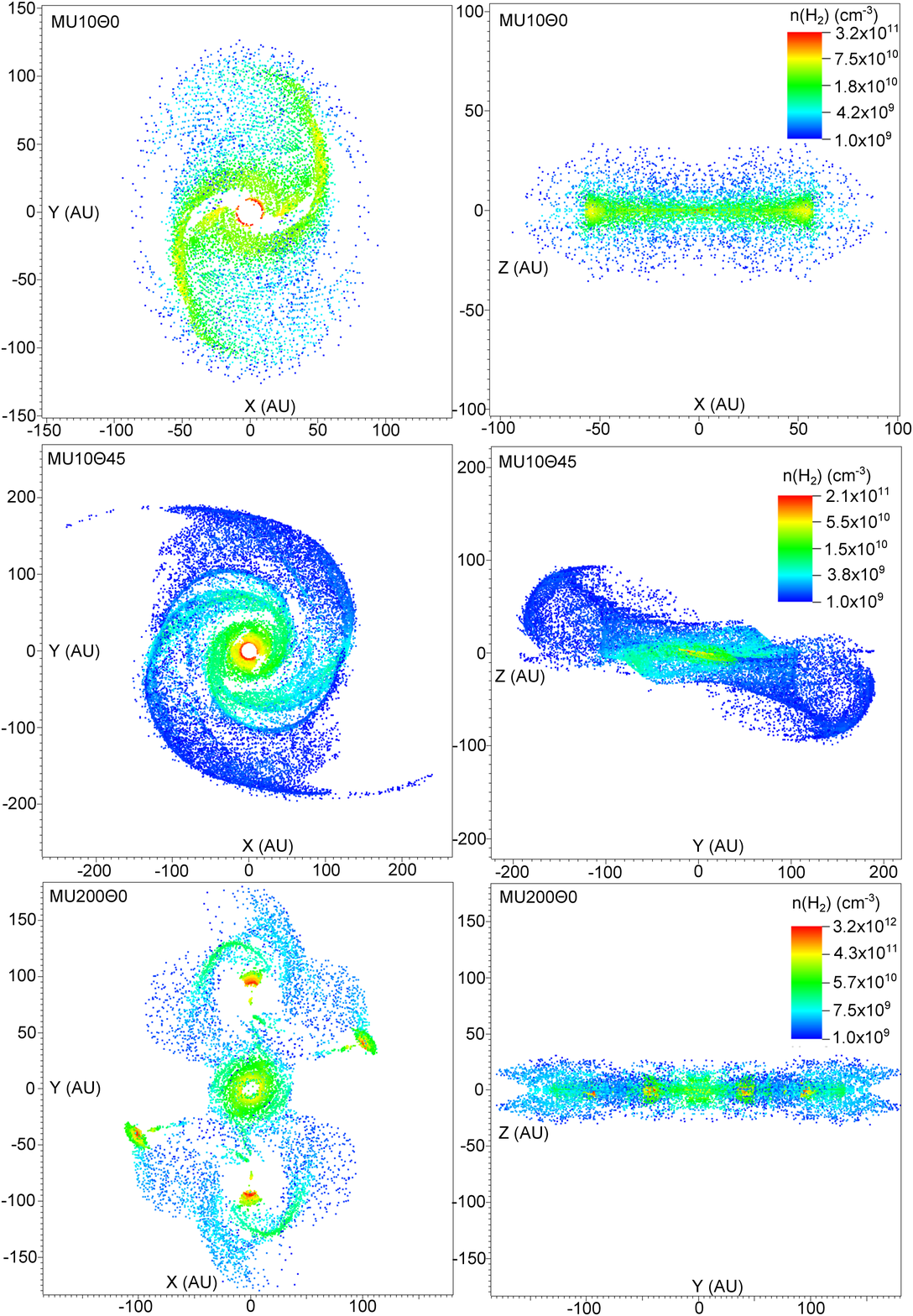}
\caption{{Density of H$_2$ (cm$^{-3}$) in the disk for the three models MU10$\Theta$0, MU10$\Theta$45 and MU200$\Theta$0.
Face-on view on the left and edge-on view on the right.
Particles are projected in the plane (xOy), (xOz) or (yOz).
{The color coding} is logarithmic.}}
\label{fig:cartes_n}
\end{figure}

\begin{table*}
\begin{center}
\caption{{Size (diameter and thickness), temperature and density ranges in the disks for models MU10$\Theta$0, MU10$\Theta$45 and MU200$\Theta$0 at the end of simulations.}\label{tab:model3D_measurements}}
\begin{tabular}{cccccc}
\tableline
\tableline
Model & Diameter (AU) & Thickness (AU) & Temperature (K)              & Density (cm$^{-3}$) & Final time (yr) \\
      &               &                & min. - max.\tablenotemark{a} & min. - max.         &                 \\
\tableline
MU10$\Theta$0  & $\sim$ 200 & 60 & 11 - 119 & 1(9)\tablenotemark{b} - 3(11) & $3.7\times 10^4$ \\
MU10$\Theta$45 & 400 & 100 & 12 - 217 & 1(9) - 2(11) & $3.9\times 10^4$ \\
MU200$\Theta$0\tablenotemark{c} & 80 & 40 & 11 - 160 & 1(9) - 3(12) & $3.9\times 10^4$ \\
\tableline
\end{tabular}
\tablenotetext{1}{min. and max. mean respectively minimum and maximum among the particles}
\tablenotetext{2}{$x$($y$) means $x\times 10^y$}
\tablenotetext{3}{values correspond to the central disk}
\end{center}
\end{table*}

\subsection{Chemical composition of the disks for the three physical models}

\subsubsection{Chemical evolution of the matter from the cloud to the disk}
\label{subsubsec:cloud_to_disk}

{Fig.~\ref{fig:sum_ice_gas} shows the sum of the abundances computed in the gas phase and at the surface of the grains (ices) compared to the total hydrogen (proton) density for a selection of species, at two different steps of the simulations: 1)~the initial condition, i.e. the composition of the parent molecular cloud, before the beginning of the collapse (namely the sphere in section~\ref{subsec:Nautilus} and \ref{interface}), and 2)~the final condition, i.e. the composition in the disk.
A change in this total abundance compared to the chemical content of the cloud implies that some processes of formation or destruction have occurred during the collapse in addition to just adsorption and desorption processes.
As the modeled object is embedded in a dense envelope, the effects of ionization and dissociation by the external UV photons are low compared to bimolecular chemical reactions.
The main result of our simulations is that the chemical content is unchanged during the collapse: there is no significant evolution of the global (ice plus gas) abundances of the species.
For example, the abundances of water {(H$_2$O)}, methane {(CH$_4$)}, carbon monoxyde {(CO)} and ammonia {(NH$_3$)} are the same in the parent cloud and the disk.\\
There are a few exceptions: HNC, CO$_2$ and HNO. HNC is very sensitive to the evolution of temperature and density during the collapse: in the case of model MU200$\Theta$0, its mean total abundance has decreased from about $5\times 10^{-7}$ (cloud composition) to $4\times 10^{-9}$ (disk composition), i.e. two orders of magnitude. The matter close to the fragments is hot so that HNC desorbs from grain and is efficiently destroyed in the gas phase by $\rm H_2COH^+ + HNC \longrightarrow HCNH^+ + H_2CO^($\footnote{\samepage
$\rm H_2COH^+ + HNC \longrightarrow HCNH^+ + H_2CO $ is a ion-dipole reaction.
New predictions for chemical rate coefficients for ion-molecule reactions have been performed by \cite{2009ApJS..185..273W} using quantum chemical calculations.
The rate coefficient formula for this kind of reaction is then different from the Arrhenius-Kooij form used up to now (see \cite{2010SSRv..156...13W} for details).
The new rate coefficient is roughly two times lower than the previous one for the temperature range $\rm [10;200]~K$ (see KIDA database), but $\rm H_2COH^+ + HNC \longrightarrow HCNH^+ + H_2CO $ remains efficient enough to destroy HNC in the gas phase in our simulations.}$^)$.
Then the HCN/HNC ratio (taking into account gas and ices) can exceed 100.
One notes on Fig.~\ref{fig:sum_ice_gas} that, even during the cloud stage, the ratio HCN/HNC is not equal to 1, but roughly equal to 10.
While the gas phase ratio is equal to 1, HCN and HNC abundances are higher on grain surfaces than in the gas phase, so that the grain surface ratio leads to a total ratio (gas and ices) of 10.
On the grain surface, HCN and HNC are photodissociated into CN and H.
Then, CN and H react on the surface, but form preferentially HCN and not HNC.
HNC can be formed on the surface using different pathways (C + NH, C + NH$_2$ or CH + NH), but these reactions are less efficient than $\rm CN + H \longrightarrow HCN$ so that the ratio HCN/HNC on grain surfaces reaches a value of 10.
One notes that as a consequence, desorption of ices could enhance the gas phase ratio HCN/HNC to a value of 10.
On the contrary but to a lesser extent, the mean global abundance of CO$_2$ has increased from the cloud to the disk, by a factor of two in the model MU10$\Theta$45. High temperature promotes CO$_2$ formation on the grain surface, particularly through $\rm OH + CO \longrightarrow CO_2 + H$ \citep{2001MNRAS.324.1054R}.
Desorption of HNO ices is followed by gas-phase reactions mainly involving CH$_3$, O, CH$_2$, and CN, that produce NO and respectively CH$_4$, OH, CH$_3$ and HCN.}

\begin{figure}
\includegraphics[width=1.0\linewidth]{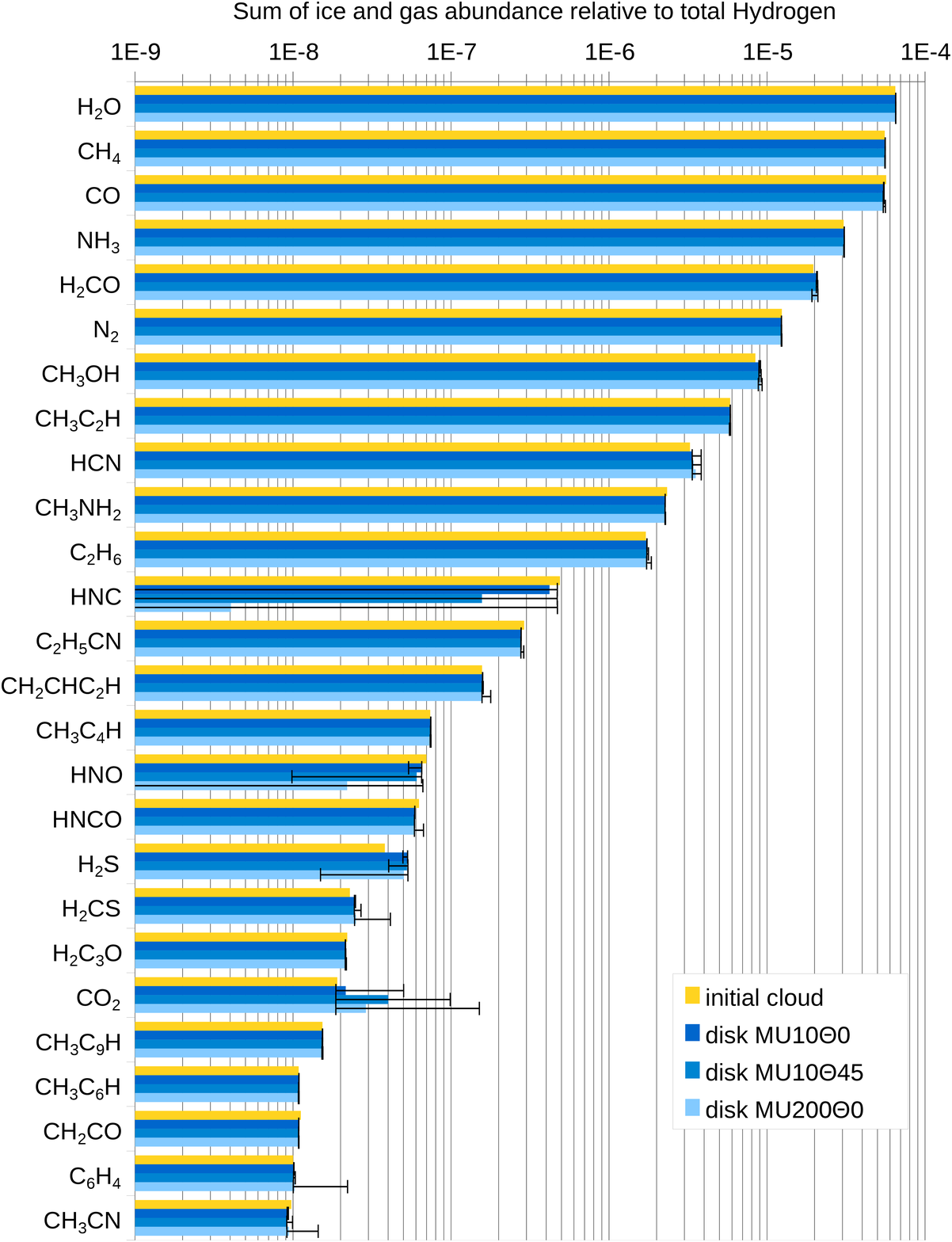}
\caption{{Abundance (in the ices and in the gas-phase) relative to total hydrogen of the most abundant species (with an abundance more than $\simeq 10^{-8}$), in the initial cloud (yellow), and in the rotating disk at the end of the simulations for the three different models: MU10$\Theta$0 (dark blue), MU10$\Theta$45 (blue) and MU200$\Theta$0 (light blue).
The plotted values for disks are the logarithm of abundance, averaged on the number of selected particles that belong to the disk.
The line segments show minimum and maximum values of species abundances.}}
\label{fig:sum_ice_gas}
\end{figure}

\subsubsection{Ice composition in the disk of model MU10$\Theta$0}

{ To illustrate the chemical composition of the ices in the disks, we present in this section the distribution of the ice composition for model MU10$\Theta$0 and for a selection of molecules observed in comets. Fig.~\ref{ab} shows the maximum (in blue) and minimum (in black) relative abundances compared to H$_2$O on the grain surfaces for these species together with the abundances observed in comets. 
Abundances below $10^{-3}$\% are not shown.
In some cases, the abundances are constant, {like} for NH$_3$ so that the minimum overlaps the maximum and only black is seen.
On the contrary, if the species abundance shows a gradient in the disk and that the smallest abundance is below $10^{-3}$, only the maximum value, in blue, appears in Fig.~\ref{ab}. 
As an example, the abundance of methanol ice (CH$_3$OH) compared to water ice is about 14~\% all over the radius of the disk (in the figure, the black band (minimum) overlays the blue one (maximum)).
The abundance of carbon monoxyde ice (CO) is high from the outer edge of the disk to around 30~AU from the central first Larson core (the blue band reaches 84~\%).
Within this radius, the CO desorbs from the grain and the ice abundance falls down to $10^{-11}$~\% (this value, being under 0.001~\%, is not visible in Fig.~\ref{ab}).
The abundance of formic acid ice (HCOOH) is low ($2\times 10^{-4}$~\%) all over the radius of the disk, which is below our 0.001~\% threshold, hence not visible in Fig.~\ref{ab}).

\begin{figure}
\includegraphics[width=1.0\linewidth]{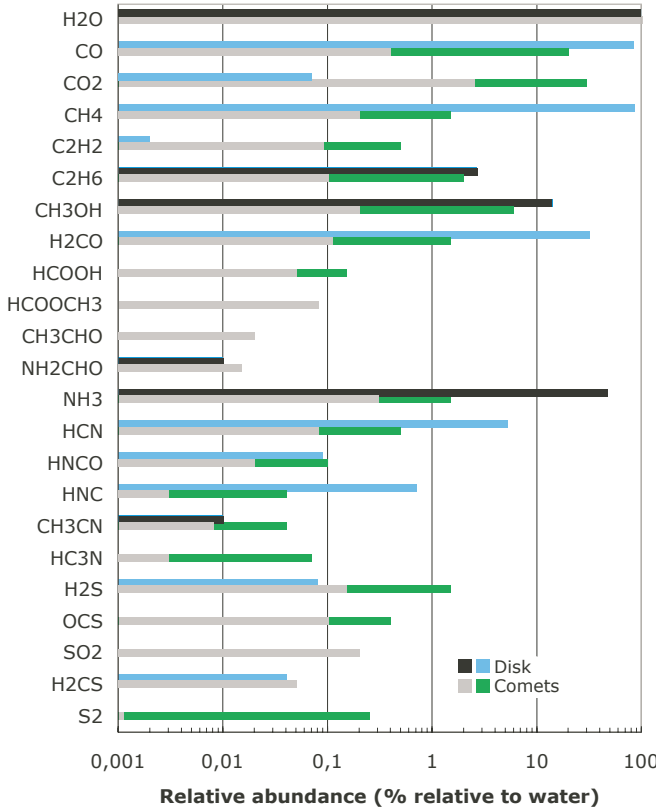}
\caption{Abundances of species in the ices relative to water on the grain surface.
The maximum and minimum computed values in the disk of model MU10$\Theta$0 are respectively shown in blue and black.
The range of measured values in comets is shown in the green portions; {in this case} the gray indicates the minimum{; if species have been observed only once, there is no green portion} (from \cite{2011ARA&A..49..471M}).
The abundance of water on grain surfaces relative to total hydrogen is $6.5\times 10^{-5}$.\label{ab}}
\end{figure}

Species listed in Fig.~\ref{ab} can be divided in three groups (Table \ref{groups}). The first group contains molecules with nearly uniform abundances as a function of the radius {(such as water ice because of its relatively high evaporation temperature)}.
In the second group, molecules are more abundant on ices in the outer {($>$25~AU)}, cold {($\sim$25~K)} disk {(such as CO ice)}, while the reverse is true for molecules in the third group {(such as C$_2$H$_2$)}.
The reason for the abundance decrease in the inner parts for the second group is evaporation due to higher temperatures.
In that case, the gas phase abundance essentially reflects the initial content of the ices, with mainly one exception, HNC, which is quickly destroyed by gas-phase reactions (see section~\ref{subsubsec:cloud_to_disk}) when the temperature is larger than $\sim$50~K (within a radius of $\sim$12 AU from the center).

\begin{table}[h]
\begin{center}
\caption{Groups of molecules {on grain surfaces} according to their behavior in the disk for model MU10$\Theta$0.\label{groups}}
\begin{tabular}{ll}
\tableline
\tableline
Homogeneous & H$_2$O, {C$_2$H$_6$}, {CH$_3$OH}, HCOOH, \\
 & {NH$_2$CHO}, NH$_3$, {CH$_3$CN} \\
\tableline
Outer disk & {CH$_4$}, {H$_2$CO}, {HCN}, {HNCO}, {H$_2$CS}, {HNC}, \\
 & {H$_2$S}, {CO}, CH$_3$CHO, S$_2$, CO$_2$ \\
 \tableline
Inner disk & C$_2$H$_2$, HCOOCH$_3$,HC$_3$N, OCS, SO$_2$ \\
 \tableline
\end{tabular}
\end{center}
\end{table}

The projected {H$_2$O$\rm _{(ice)}$, CO$\rm _{(ice)}$ and C$_2$H$_{2{\rm (ice)}}$ abundances} (relative to total hydrogen) computed in the disk {for the model MU10$\Theta$0} are shown in {Fig.~\ref{fig:ab_glace_mu10theta0}}.
{These three molecules illustrate the different behaviors presented in Table~\ref{groups}.}
The temperature is rather low, below 20~K in the majority of the disk (from the outer edge to $\sim$40~AU) whereas the total H density presents {variations between $10^9$ and $8\times 10^{10}$~cm$^{-3}$ from the outer edge to $\sim$15~AU}.
The inner regions (within a radius $r\approx 20$~AU) present the warmest part of the disk with temperatures up to about {120~K}. The abundance of water (relative to total hydrogen) trapped on interstellar grains is constant across the disk at a value of $6.5\times 10^{-5}$. The decrease of the CO ice abundance is a direct consequence of thermal evaporation in this region.
{When the temperature increases (as a function of the radius), electronic recombination reactions in the gas phase enhance the abundance of C$_2$H$_2$.
The gas-phase C$_2$H$_2$ is then adsorbed on grains.
Desorption is not efficient for this species because its adsorption energy is high enough.}

\begin{figure}
\includegraphics[width=1.0\linewidth]{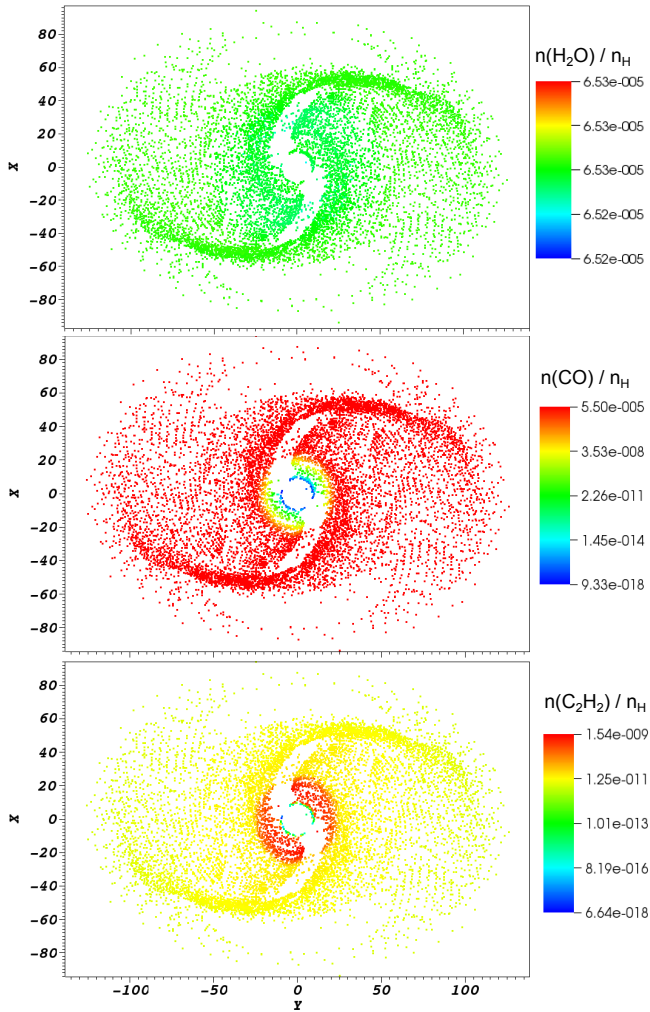}
\caption{{H$_2$O, CO and C$_2$H$_2$ abundances relative to total H in the ices in the rotating disk of model MU10$\Theta$0.
The ''central and spiral hole pattern'' is composed of particles that do not respect the criteria of a disk according to \cite{2012A&A...543A.128J} (i. e. belong to the core or are not supported by rotation).}}
\label{fig:ab_glace_mu10theta0}
\end{figure}

Results for model MU10$\Theta$45 and MU200$\Theta$0 are qualitatively similar to those of MU10$\Theta$0, with some different distributions of molecules according to the local temperature and density conditions.
Due to fragmentation and the presence of four hot spots in the model MU200$\Theta$0, ices are sublimated in these hotter regions, and molecules are eventually destroyed in the gas phase.
The higher temperature of the matter next to these regions lead to a more active surface chemistry which enhances the formation of complex organic molecules such as the methylformate $\rm HCOOCH_3$.
The disk of the model MU10$\Theta$45 is on average a little bit warmer, which leads to a smaller abundance of ices but a higher abundance in the gas phase so that the total abundance is globally not changed in a significant amount (see Fig.\ref{fig:sum_ice_gas}).

\section{Discussion}

\subsection{Comparison with comets}

Our model stops just before the second collapse and the formation of the second Larson core, which is early in the star formation process (see section~\ref{sec:limits}).
However, recent hydrodynamic simulations \citep{2010ApJ...724.1006M,2011MNRAS.413.2767M,2010MNRAS.404L..79B,2011MNRAS.417.2036B} have shown that the central part (radius $\lesssim 1$~AU) of the first Larson core will form the second Larson core, that is to say the protostar, whereas the outer part of the first Larson core will evolve in a disk.
During the protostellar phase, a small fraction of the first core material should have been transported to the outer region of the disk, and not been accreted to the protostar (see discussion of FA12).\\
{Considering the partial interstellar origin of the molecules observed in comets \citep{2000A&A...353.1101B,2011ARA&A..49..471M}, it is then interesting to compare the comet chemical composition and the results of our models.}
Due to the fact that our disk is very young, the computed ice can be considered as "pre-cometary", and so this comparison must stay qualitative.

{The ice abundances of C$_2$H$_6$, CH$_3$OH, NH$_2$CHO and CH$_3$CN found in our disk for model MU10$\Theta$0 are within a factor of 2 of those found in comets (see  Fig.~\ref{ab}). It is also the case for HNCO, H$_2$S and H$_2$CS in the outer part of the disk where the temperature is low enough to maintain these species on grain surface. These species represent about one third of the observed species in comets.}
On the other hand, CO$_2$, C$_2$H$_2$, HCOOH, HCOOCH$_3$, CH$_3$CHO, HC$_3$N, OCS, SO$_2$ and S$_2$ are strongly underproduced in our model, { in the entire disk}, compared to comets.
Carbon dioxide (CO$_2$) and carbon sulfide (OCS) are known to be abundant on interstellar ices \citep{1997ApJ...479..839P,2000ApJ...536..347G} but not much produced by astrochemical models at low temperature because carbon, oxygen and sulfur only diffuse slowly on surfaces \citep{2001MNRAS.324.1054R}.
For example, the surface reaction $\rm HCO + O \longrightarrow CO_2 + H$ is the main pathway in our model for CO$_2$ production, but is in competition with $\rm HCO + H \longrightarrow H_2CO$ which is faster.
In our results, CH$_4$ and NH$_3$ are reservoirs of carbon and nitrogen whereas they are minor species in comets.
The methane is probably overproduced in our model because we underestimate the formation of CO$_2$: the available carbon on grain surfaces is quickly hydrogenated so the carbon is rather locked into methane.
For nitrogen, the reservoir is not known in comets \citep{2004come.book..391B} and we probably lack some mechanisms to form other N-bearing species \citep{2012PNAS..10910233D}.
Finally, HCOOH, CH$_3$CHO and HCOOCH$_3$ are known to trace warmer regions where radicals can move at the surface of the grains, and are typically observed in warmer star forming regions \citep{2009ARA&A..47..427H}.

Comets seem to be made of molecules formed in different physical conditions.
{Table~\ref{tab:formation_place} illustrates this idea proposing a formation place of molecules where the required physical conditions are present.}
The HNC molecule for instance, is a strong signature of cold interstellar like-chemistry. Other molecules on the contrary, such as HCOOCH$_3$, require
temperatures of about 40 to 60~K to be formed on surfaces
\citep{2006A&A...457..927G}, and as a consequence are not produced by our model
in the disk in sufficient amount. This mixed composition may be a strong
indication of radial mixing in the protoplanetary disk. This process is not included in our simulations.
In fact, we find that the composition of ices are almost constant in the outer regions but that there is a gradient of abundances within 40 AU, a size consistent with the assumed region of formation of comets in our solar system (see \cite{2011ARA&A..49..471M} and references within it).

\begin{table}[h]
\begin{center}
\caption{Possible formation place of molecules observed in comets.\label{tab:formation_place}}
\begin{tabular}{ll}
\tableline
\tableline
Cloud & C$_2$H$_6$, CH$_3$OH, HNCO, \\
(low temperature) & NH$_2$CHO, CH$_3$CN, H$_2$S, H$_2$CS \\
\tableline
Cloud & CH$_4$, H$_2$CO, HCN, \\
+ desorption process\tablenotemark{a} & HNC, CO \\
 \tableline
Close to the first core & CO$_2$, C$_2$H$_2$, CH$_3$CHO, OCS, \\
(warmer region) & HCOOH, HCOOCH$_3$ \\
\tableline
\end{tabular}
\tablenotetext{1}{Ices formed in the cloud excessively compared to comet composition, that are lowered by desorption process when reaching the disk.}
\end{center}
\end{table}

\subsection{Impact of initial conditions on the survival of molecules during the collapse}

In order to assess the sensitivity of our results to the composition of
the initial dense core, we computed the chemical composition for three different particles of model MU10$\Theta$0 using several initial conditions. The three particles reach at the end
of the simulation respectively the cold region ($\sim 10$~K), the warm region
($\sim 30$~K) and the hot region ($\sim 100$~K) of the disk.

\subsubsection{A denser molecular cloud}

As a first test, we constructed our initial conditions from a Nautilus run with
a density of $2 \times 10^6$ cm$^{-3}$, i.e. ten times higher than the one of our
standard initial conditions described in section~2.2, for $6\times 10^5$~yr. The resulting dense core composition is in fact
similar to the standard one. Differences mostly concern atoms, whose abundances are lowered by one
to two orders of magnitude at the final time. However, the reservoirs of carbon, oxygen and nitrogen
remain CO, CO$_2$, H$_2$O and NH$_3$ and their abundances are not changed
significantly.\\
Consequently, the evolution of our three particles during the subsequent
collapse keeps the properties described in section
\ref{subsubsec:cloud_to_disk}: the most abundant molecules fully survive, only
trace species are altered, the most sensitive being HNC, with an abundance
lowered by 6 orders of magnitude for the particle reaching the hot region of
the disk.

\subsubsection{An older molecular cloud}

Knowing the large uncertainties pertaining to the age of molecular clouds (see for example \cite{2001ApJ...562..852H}, \cite{2006ApJ...646.1043M}, \cite{2007RMxAA..43..123B}, and \cite{2011ApJ...739L..35P}), we
constructed another set of initial conditions for the same physical parameters
as for the standard one, but for an
evolution during $6\times 10^6$ yr, i.e. ten times longer. The resulting composition is different in such a case. The most
notable difference is that CO is no longer a carbon and oxygen reservoir.
The abundance of CO in the cloud is decreased in favor of H$_2$O and CH$_4$:
chemistry being out of equilibrium, the longer integration time allows
adsorption of CO on the grain surface, which is followed by successive
hydrogenation that leads to CH$_3$OH formation. Then, photodissociation of
CH$_3$OH by cosmic ray induced UV photons produces CH$_3$ and OH, which are
hydrogenated to form CH$_4$ and H$_2$O \citep[see][for details about this process]{2007A&A...467.1103G}.\\
With such initial conditions, our main conclusions remain unchanged, only trace
species are altered. Interestingly enough, CO is not a reservoir in this case
and shows a significant evolution during the collapse, with an abundance
decreasing by up to 3 orders of magnitude for the particle arriving in the warm
region of the disk, because of adsorption followed by grain surface reactions.
This effect is less strong for the other two particles: for the one arriving in the hot
zone, CO desorbs sooner, while the one arriving in the outer parts sees lower
temperatures, both effects making grain surface reactions less efficient.

\subsubsection{Observed abundances in dense clouds as initial conditions}

Another natural choice for the initial conditions is to use the composition
actually observed in dense cores. The gas phase abundances were taken from
observations of TMC-1(CP)\footnote{\samepage A compilation of gas phase
molecular abundances observed in this source will be given by Ag{\'u}ndez \&
Wakelam, submitted to chemical review. Values used in the current study come
from 
\cite{1985ApJ...290..609M}, \cite{1987ApJ...315..646M},
\cite{1989ApJ...345L..63M}, \cite{1991A&A...247..487S},
\cite{1992ApJ...386L..51K}, \cite{1992ApJ...396L..49K},
\cite{1992IAUS..150..171O}, \cite{1993A&A...268..212G},
\cite{1994ApJ...422..621M}, \cite{1994ApJ...427L..51O},
\cite{1997ApJ...486..862P}, \cite{1998A&A...335L...1G},
\cite{1998FaDi..109..205O}, \cite{1998A&A...329.1156T},
\cite{1999ApJ...518..740B}, \cite{2000ApJ...539L.101S},
\cite{2001ApJ...552..168F}, \cite{2003A&A...402L..77P},
\cite{2006ApJ...643L..37R}, \cite{2006ApJ...647..412S},
\cite{2008A&A...478L..19A}, \cite{2009ApJ...690L..27M}, and
\cite{2011A&A...531A.103C}.  }.  
The ice abundances of CO$_2$, CO,
NH$_3$, CH$_3$OH, and OCS were taken from observation of Elias 16, a K1 giant behind
TMC, considered as a quiescent environment \citep{2004ApJS..151...35G}. The
abundance of water ice is set to its median value observed in cold dense 
sources, which is about $5\times 10^{-5}$ relative to total hydrogen
\citep{2004ApJS..151...35G,2004A&A...426..925P,2004ASPC..309..547B,2011ApJ...740..109O}. To conserve the same elemental abundances as in the standard model, the atomic abundances have been adjusted. \\
The chemical evolution during the collapse is more significant in this case.
Apart from the five most abundant molecules, CO, H$_2$O, CO$_2$, NH$_3$, and
CH$_3$OH, who survive the collapse and remain reservoirs, most species see
their abundances change during the collapse. This is due to the fact that the
initial condition is not a quasi-stationary state for our chemical network, as
our network does not reproduce exactly the observed abundances in dense clouds.
Consequently, as soon as the calculation starts, rapid gas phase reactions tend
to relax the system towards a chemical state more consistent with our chemical
network and most abundances change. \\ However, it is worth noting that, even
in this case, the most abundant species survive the collapse.\\

This study of the influence of initial conditions strengthens the
conclusions of section \ref{subsubsec:cloud_to_disk}: the memory of the
composition of the prestellar core is kept throughout the first phases of the
formation of the disk, at least for the most abundant species.
 
\subsection{Comparison with other three-dimensional numerical simulations}

{FA12 present three-dimensional modeling results of a collapsing dense core until the formation of the first hydrostatic core, that is to say the first Larson core.
They use a similar interface as ours: they use fluid particles that trace the physical conditions of the matter during the collapse, and compute the chemical evolution on these particles as a post-process.
The main differences with our model are that FA12, on a physical point of view, do not compute the effect of magnetic field on the collapse process, and on a chemical point of view, use a high temperature gas-phase network (for a range of 100-800~K).
Since we focus on the disk composition in the present paper, where the temperature does not exceed 300~K, we do not expect our results to be affected by this limitation (see section~\ref{sec:limits}).}

Figure~\ref{fig:Tn_radius_mu10} shows the density and temperature of particles in the equatorial plane (area where $|z|<10$~AU) as a function of the distance from the central core, for the model MU10$\Theta$0.
At a given radius, the range of values come from the asymmetry of the collapsing core: in particularly, because of the presence of spiral arms.
The figure can be compared to panel (b) of Figure 9 of FA12 (although the integration time is not exactly the same): the temperature and density profiles are qualitatively similar, except within 1-2~AU where the density and temperature are higher in the case of FA12. 
In our model, an additional support against gravitational collapse exists, magnetic support, which limits contraction.
As a consequence, the inner computed temperature and density are lower in our case. 
One of the main conclusion of FA12 is that the total gas and ice abundances of many species remain unaltereted during the collapsing process until the temperature reaches about 500~K, that is to say at a spherical radius $r\lesssim 3$~AU.
The gas phase abundances are determined via desorption process, depending on the adsorption energy of species on the surface of the grains.
Our results show similar conclusion, at least for the disk component: except at low radius, the chemical composition is globally unaltered during the collapse with a gas phase abundances mainly determined by desorption. In the warmer parts of the disk, that is to say at lower radius, we also observe a formation of carbon chains and large organic molecules such as C$_2$H$_2$, HC$_3$N and HCOOCH$_3$ (see Table~\ref{groups}).

\begin{figure}
\includegraphics[width=1.0\linewidth]{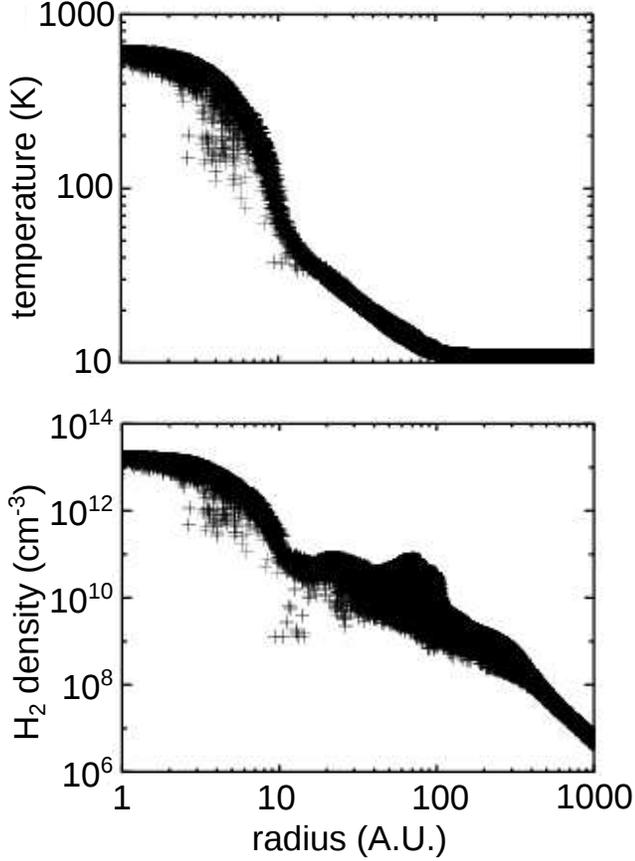}
\caption{{Temperature and density of particles that belong to the equatorial plane (where $|z|<10$~AU) as a function of the radius at the end of the simulation for the model MU10$\Theta$0.}}
\label{fig:Tn_radius_mu10}
\end{figure}

\subsection{Limits of the model}\label{sec:limits}

{The simulations from the RMHD physical model {\ttfamily{RAMSES}} stop just before the start of the second collapse.
The formation of this second core increases the temperature in the disk at larger radii as suggested by the 1D modeling of \citet{2013arXiv1307.1010V}.
Compared to our simulations, this may mainly change the radii at which molecules evaporate.
The importance of this second phase will depend on the temperature gradient obtained in the disk.
This issue will be investigated in further studies.\\
}

{The physical model {\ttfamily{RAMSES}} integrates the equations of ideal magnetohydrodynamics, assuming the medium to be a perfect conductor.
The matter is then frozen on the magnetic lines.
If we consider a medium that is not a perfect conductor, that is to say in the case of non ideal (or resistive) magnetohydrodynamics, the neutral matter will be able to cross magnetic lines.
This induces a friction between ions and neutrals, called ambipolar diffusion \citep{1956MNRAS.116..503M,1991ApJ...373..169M}.
Ambipolar diffusion generates a net velocity difference between ions and neutral that can be sufficient to overcome activation barriers, thereby generating efficent reaction pathways.
Future work would be done to couple non ideal magnetohydrodynamics with chemistry in order to take this phenomenon into account.}
Recent works have been done to integrate ambipolar diffusion in {\ttfamily{RAMSES}} \citep{2012ApJS..201...24M}.

{As mentioned in section~\ref{interface}, the chemical model Nautilus is used as a post-processing of the {\ttfamily{RAMSES}} computation.
There is no chemical feedback on the physical structure of the collapsing dense core.
But, a radiative cooling of the matter due to chemical species such as CO, CH$_4$, H$_2$O, C and O can occur.
The cooling rate of a given chemical species depends in particular on the temperature and the density of the medium, and the abundance of the species \citep[see for example][]{1993ApJ...418..263N,2011IAUS..280P.149D}.
However, at high density such as in the case of our modeling, the cooling is mostly dominated by the radiative emission of the dust.}

In our modeling, each particle is independent of the others from a chemical point of view. This is valid only if mixing effects can be neglected, both molecular diffusion and turbulent mixing.\\
For matter collapsing onto the core and the disk, our simulations do not show any turbulent motion, and in general turbulent mixing is not expected to be efficient during this phase.
The influence of molecular diffusion can be qualitatively estimated by comparing the characteristic diffusion timescale $\tau_\mathrm{diff}$ for two nearby tracer particles, with the time $\tau_\mathrm{disk}$ for these particles to enter the disk.
$\tau_\mathrm{diff}$ is given by:
\begin{equation}
\tau_\mathrm{diff} \simeq \frac{L^2}{c_s\lambda} \simeq L^2 n_{\rm H_2} \sigma \left(\frac{\mu_m m_{\rm H}}{k_B T}\right)^{1/2}
\end{equation}
where $L$ is the distance between the two particles, $c_s$ is the sound speed (typical velocity of molecules inside the lagrangian particle), $n_{\rm H_2}$ is the molecular hydrogen number density,
$\lambda$ is the mean free path of gas phase molecules and $\sigma$ is the collision cross section.
$\tau_\mathrm{disk}$ is roughly equal to the free fall time $\tau_\mathrm{ff}$:
\begin{equation}
\tau_\mathrm{disk} \simeq \tau_\mathrm{ff}=\left(\frac{3\pi}{32G\rho_0}\right)^{1/2}
\end{equation}
where $G$ is the gravitational constant.
$\tau_\mathrm{ff}$ is close to the final time given in table~\ref{tab:model3D_measurements} ($\tau_\mathrm{ff} \simeq 3\times 10^4$~yr).
In the collapsing region, $L>1$~AU, $T=10$~K, and $n_{\rm H_2}=10^6$~cm$^{-3}$, so that $\tau_\mathrm{diff}> 2\times 10^5$~yr which is higher than $\tau_\mathrm{ff}$.
Therefore particles collapse faster than they can mix with each other.\\
When the particles reach the disk, the situation changes.
Although our numerical resolution is insufficient to allow for turbulence to develop in the simulation, turbulence is expected to exist in these objects.
In such a case, particles can no longer be considered as isolated, molecules are expected to steadily move from one particle to another.
The detailed coupling of chemistry with turbulence being beyond the scope of the present paper, we considered only average disk abundances, as displayed in Fig. \ref{fig:sum_ice_gas}.
Note, however, that turbulent mixing is expected to homogenize the disk composition, so that the amplitude of our range of disk abundances may be overestimated.

{Finally, our chemical network contains reactions whose rate coefficients are evaluated from 10 to 300~K.
Above 300~K, endothermal reactions or reactions that have large activation barrier can become efficient.
Since the central part of the collapsing dense core is warmer than 300~K, a high temperature network should be used for this region.
However, the component discussed in this paper, i.e. the disk, does not experience such high temperature.
Including the high temperature gas phase network from \citet{2010ApJ...721.1570H} will be a future improvement of our modeling.}

\section{Conclusion}

In this article, we present the 3D chemical composition computed during the collapse of a dense core, up to the first Larson core, in which the physical conditions are computed by a RMHD model {and the chemical composition is computed by a full gas-grain model.}
{Our main result is that the total abundance (gas plus ices) remains mostly unchanged from the parent cloud to the disk, except for some species such as HNC, which is destroyed, and CO$_2$, which is formed.
The chemical composition in the outer part of the young protoplanetary disk reflects the initial composition of the parent molecular cloud, until the temperature in the disk reaches the evaporation temperature of the molecules.
For this reason, the abundances in the disk depend on the initial chemical composition of the simulation.
We also find some similarities between our computed ice composition in the disk and the molecular abundances found in comets (for about one third of the observed species). }
For these molecules, there is no need to invoke an in situ formation {that is to say} the abundance observed in comets can reflect the initial composition of the envelope {or the parent molecular cloud}.

Our simulations stop early in the history of the disk so that the {chemical} composition may be modified later in the proto-planetary disk. 
Future calculations will be done to check this hypothesis.

\acknowledgments

This research was partially funded by the program PCMI from CNRS/INSU.
Ugo Hincelin was funded by a grant from the french ''R\'egion Aquitaine'',
and acknowledges current postdoctoral fellowship support from the National Science Foundation through a grant to Eric Herbst.
The research of Benoit Commer\c con is supported by the ANR Retour Postdoc program and from the CNES.
Benoit Commer\c con acknowledges the postdoctoral fellowship support from the Max-Planck-Institut f\H{u}r Astronomie where part of this work was conducted.
The {\ttfamily{RAMSES }}\rm  calculations have been performed  at CEA on the DAPHPC cluster and the Nautilus calculations were performed using JADE cluster resources from GENCI-CINES (Grand Equipement National de Calcul Intensif - Centre Informatique National de l'Enseignement Sup\'erieur).
Some kinetic data we used have been downloaded from the online database KIDA (KInetic Database for Astrochemistry, \url{http://kida.obs.u-bordeaux1.fr}, \cite{2012ApJS..199...21W}).
Ugo Hincelin thanks Eric Herbst for helpful discussions on interstellar ice abundances.
The authors thank Patrick Hennebelle for giving time on JADE cluster to finish the Nautilus calculations, and Franck Selsis for helping on data management.
The authors also thank the referee for constructive remarks which helped to make this paper clearer and more useful to the reader.

\end{document}